\documentclass[aps,prd,notitlepage,showpacs,nofootinbib,superscriptaddress]{revtex4-1}

\usepackage{graphicx}
\usepackage[utf8]{inputenc} 

\usepackage{amsmath}
\usepackage{amssymb}
\usepackage{float}
\usepackage{comment}
\usepackage{slashed}
\usepackage[normalem]{ulem}

\usepackage{caption}
\usepackage{subcaption}
\usepackage{multirow}
\usepackage{rotating}

\usepackage[usenames,dvipsnames]{color}
\usepackage[colorinlistoftodos]{todonotes}
\usepackage[colorlinks=true,citecolor=darkred,urlcolor=darkred, pdfborder={0 0 0}]{hyperref}
\usepackage[normalem]{ulem}

\definecolor{darkred}{rgb}{0.6,0,0}
\usepackage[colorinlistoftodos]{todonotes}

\definecolor{linkcolor}{rgb}{0,0,0.5}
 
\newcommand {\ignore}[1]{}

\def\gsim{\raise0.3ex\hbox{$\;>$\kern-0.75em\raise-1.1ex\hbox{$\sim\;$}}}
\def\lsim{\raise0.3ex\hbox{$\;<$\kern-0.75em\raise-1.1ex\hbox{$\sim\;$}}}

\providecommand{\be}{ \begin{equation} } 
\providecommand{\ee}{ \end{equation} }
\providecommand{\bea}{\begin{eqnarray}}
\providecommand{\eea}{\end{eqnarray}}

\providecommand{\to}{\rightarrow}

\usepackage{soul}

\definecolor{mightnightblue}{RGB}{25,25,112}

\definecolor{brown}{rgb}{0.59, 0.29, 0.0}

\newcommand {\black} {\color{black}}

\def\21{$\mathrm{SU(2)_L \otimes U(1)_Y}$}

\newcommand{\AddrUCOL}{Facultad de Ciencias, Universidad de Colima, Colima 28010, Mexico.}
\newcommand{\AddrDUAL}{Dual CP Institute of High Energy Physics, C.P. 28045, Colima, México }
\newcommand{\AddrUCN}{Departamento de Física, Universidad Católica del Norte, \\
Avenida Angamos 0610, Casilla 1280, Antofagasta, Chile.}

\begin{document}

\title{{\color{violet}Sub-GeV $\boldsymbol{U(1)_{R}}$ gauge boson to address the proton radius discrepancy}}

\author{Carlos Alvarado} \email{calvara@dcpihep.com}
\affiliation{\AddrDUAL}
\author{Alfredo Aranda} \email{fefo@ucol.mx}
\affiliation{\AddrUCOL}
\affiliation{\AddrDUAL}
\author{Cesar Bonilla} \email{cesar.bonilla@ucn.cl}
\affiliation{\AddrUCN}
\affiliation{\AddrDUAL}

\begin{abstract}
\vspace{0.5cm}
We propose a Standard Model extension by a $U(1)_{R}$ gauge symmetry where only right-handed chiral fermions can carry a non-trivial charge. Here we show that the simplest anomaly-free solution to accommodate the proton charge radius discrepancy takes right-handed muons $\mu_R$ and first generation quarks, $u_R$ and $d_R$. Consistency with the latest muon's $(g-2)$ measurements is achieved through an extra light scalar, which itself must lie in the tens of MeV mass range to be viable.
\end{abstract}

\maketitle

\section{Introduction}
\label{sec:intro}

The Standard Model (SM) of particle physics successfully describes three of the four known fundamental interactions of nature: electromagnetic,  weak, and strong. The tremendous accuracy of the SM has been corroborated by contemporary collider experiments at high energies such as the Large Hadron Collider (LHC). Nevertheless, this and other experiments have found several subtle deviations from the SM predictions that remain to be accounted for. For some of these anomalies, if they persist and get confirmed,  the presence of new physics is required.
 
Some of those deviations have received particular attention during the last decade with revitalized interest provided by new measurements. It is peculiar that they all somehow involve the muon, suggesting that subtle effects in its interactions might a be key to new physics. Examples are the $B$-meson anomalies, which are deviations from the SM predictions in angular observables and the ratios $R_{K(K^{*})}=\mathcal{B}r(B\to K^{(*)}\mu^{-}\mu^{+})/\mathcal{B}r(B\to K^{(*)}e^{-}e^{+})$ that might point to violation of lepton-flavor universality. These have captured a lot of theoretical interest with plenty of effort channeled towards setups that extend the SM with additional scalars and/or gauge bosons coupled to quarks and leptons in a non-universal manner \cite{Bonilla:2017lsq}. As of today, the deviations persist at a level above $3\sigma$, based on the latest results from the LHCb collaboration at the LHC \cite{LHCb:2021trn,Altmannshofer:2021qrr}. Another example of anomalies is the latest measurements of the muon's $(g-2)$ \cite{Abi:2021gix}, for which attempts to accommodate it within SM extensions containing \textit{muonic} mediators already comprises a vast literature.

Our focus is in yet another instance of tentative new physics involving the muon. For decades, the study of the structure of the hadrons found in nuclei \textemdash protons and neutrons\textemdash has improved as a result of  increasingly refined techniques such as lepton-nucleon scattering and the spectroscopy of hydrogen-like atoms. Among the long-standing studied observables there is the (root-mean-square) proton charge radius, $r_{p}$. Despite the good agreement between measurements of $r_{p}$ by experiments involving electrons prior to 2010, in that year the CREMA collaboration found a highly discrepant $r_{p}$ value through spectroscopy of hydrogen-like atoms where the muon replaces the electron (muonic hydrogen) \cite{Pohl:2010zza}. The muonic result improved in 2013 \cite{Antognini417} and continued to be discrepant with the electronic-measurement averages compiled and published by the CODATA committee in 2014 \cite{Mohr:2015ccw}
\begin{align}
r_{p}^{(e)}   &= 0.8751 \pm0.0061 \text{ fm}\notag \\
r_{p}^{(\mu)} &= 0.84087\pm0.00039 \text{ fm}~. \label{eq:rpeANDmu}
\end{align}
Fast forward to 2021, several experiments with electrons and form-factor analyses of the data have found a degree of consistency with the muonic $r_{p}$ value. In particular the CODATA 2018 average \cite{RevModPhys.93.025010} and posterior reports \cite{Bezginov1007} tend to favor the muonic $r_{p}$ value. Yet, since a few other recent experimental collaborations \cite{PhysRevLett.120.183001,Mihovilovic:2019jiz} have reported $r_{p}$ values that persistently agree with the CODATA 2014 $r_{p}$ value, the proton charge radius puzzle is certainly ameliorated but not conclusively gone\footnote{Two comprehensive reviews on the proton charge radius are found in Refs. \cite{Carlson:2015jba} and \cite{Gao:2021sml}.} \cite{PhysRevLett.120.183001,Mihovilovic:2019jiz}. Similarly, given that the value found by Grinin et. al. \cite{Grinin1061} is offset the muonic value by about two standard deviations, the puzzle is kept alive.

With the ongoing progress in higher order theoretical corrections in scattering and spectroscopy and the projected sensitivity of near-future experiments, the community has also entertained the possibility that these $r_{p}$ discrepancies are a hint of new physics consisting on muonphilic interactions of the proton \cite{Carlson:2012pc,Batell:2011qq,Perelstein:2020suc,Zhu:2021vlz}. This is the approach taken in this paper: we follow the hypothesis that the $\Delta r_{p}^{2}$ discrepancy between the two values in Eq. (\ref{eq:rpeANDmu}) will persist, and is addressed by new degrees of freedom of a complete, anomaly-free gauge $U(1)$ model. Our setup is also consistent with other relevant constraints, including LHC bounds on new scalars and those from the latest muon $(g-2)$ results by Fermilab \cite{Abi:2021gix,Albahri:2021kmg} and the Lattice QCD group \cite{Borsanyi:2020mff}.

Our work is organized as follows: the new gauge model and its field content is described in in Sec. \ref{sec:content}, with the proton charge radius discrepancy reviewed and addresssed in Sec. \ref{sec:radius}. Contributions to the muon's $(g-2)$ are discussed in Sec. \ref{sec:gminus2}, where we also study numerically the parameter space where $r_{p}$ can be accomodated while consistent with collider limits on scalars and the muon's $(g-2)$. Closing remarks follow afterwards.

\section{Model selection}

\label{sec:content}

In this scenario the SM gauge group is enlarged by an Abelian gauge $U(1)_R$ symmetry (and its corresponding $Z'$ gauge boson) where only right-handed(RH) chiral fermions may have non-trivial transformation under the new symmetry. Specifically, only first generation RH quarks and the RH muon are charged under the new $U(1)_R$ gauge symmetry\footnote{Other examples and uses of Abelian gauge extensions can be found in Refs. \cite{FAYET1990743,Fayet:2016nyc}.}.

In addition, the model contains the following $U(1)_{R}$-charged scalars, needed to break $U(1)_{R}$ spontaneously and give a mass to the $Z'$: a complex SM singlet $s$, and a $SU(2)_{L}$ doublet $\Phi_2$ (same hypercharge as the SM Higgs field, denoted as $\Phi_{1}$) which will also be responsible for generating the masses of the muon, up, and down quarks. As it is shown in the next section, due to its relative size to the vacuum expectation value (vev) of $\Phi_{2}$, the vev of $s$ generates most of the $Z'$ mass while accomodating the charge radius discrepancy. Meanwhile, a range of sizes for the $\Phi_{2}$ vev is discussed at the end of this section. The rest of the SM particles are uncharged under the new symmetry, including the Higgs-like doublet.

The fact that SM fermions of different generations carry distinct $U(1)_{R}$ quantum numbers implies that gauge anomalies do not cancel automatically per generation, thus extra (chiral) fermion content is needed to satisfy the anomaly-free conditions. For this purpose one SM singlet RH neutrino $\nu_R$ is introduced into the model (with a yet unspecified charge $q_{\nu}\neq0$ that does the job when exactly one quark and one lepton generations are charged\footnote{$\nu_{R}$ is not tied to the any lepton generation.}).

\begin{figure}[h!]
\centering
\includegraphics[scale=0.6]{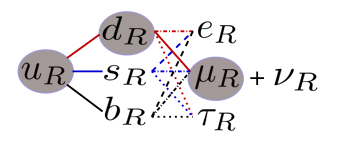}
\caption{Set of right-handed fermion combinations free of gauge anomalies. The solid red line connecting the three blobs illustrates the link between nucleons and muons as demanded by the proton radius puzzle.}
\label{fig:afsols}
\end{figure}

Given the introduction of the SM singlet $\nu_R$, there are 27 different anomaly-free scenarios that one could generate. The diagram in Fig.~\ref{fig:afsols} exemplifies the 9 gauge anomaly-free combinations involving the RH $u$-quark, with all three RH $d$-type quarks and RH charged leptons. The other 18 solutions involve $c_R$ and $t_R$. Seeking to establish a connection between nucleons and muons that tackles the proton radius puzzle, the valence RH-quarks of the first generation and the second generation lepton  are the ones given a non-trivial charge under $U(1)_{R}$. This choice is illustrated by the solid red line in Fig. \ref{fig:afsols}. For reference, Table \ref{tab:content} displays the rest of anomaly-free charge assignments of the model's matter content in terms of an arbitrary charge $q_{\nu}$. 

\setlength\tabcolsep{8pt} 
\begin{table}[H]
\centering
\begin{tabular}{|c|ccccccccc|ccc|} 
\hline
            & $Q_{L_{(1,2,3)}}$ & $u_{R}$    & $(c_{R},t_R)$    & $d_{R}$ & $(s_{R},b_R)$ &  $L_{(1,2,3)}$ & $(e_{R},\tau_R)$ &  $\mu_{R}$ &$\nu_R$  & $\Phi_1$  & $\Phi_2$&  $s$  \\
\hline
$SU(2)_{L}$ & ${\bf2}$ & ${\bf1}$    & ${\bf1}$  & ${\bf1}$    & ${\bf1}$  & ${\bf2}$ & ${\bf1}$  & ${\bf1}$    & ${\bf1}$ & ${\bf2}$ & ${\bf2}$ &    ${\bf1}$  \\     
\hline
$U(1)_R$    & 0        & $q_\nu$  &0        &   $-q_\nu$  &0    & 0 &0     & $-q_\nu$    & $q_\nu$  & 0  & $q_\nu$ &    $q_s$    \\  
\hline
\end{tabular}\caption{Particles with nonzero gauge $U(1)_{R}$ charge assignments.}
\label{tab:content}
\end{table}
We choose to work out the $q_{\nu}=+1$ solution for the sake of generality. In this scenario, a Dirac neutrino mass is generated through the Yukawa term $\sim\overline{L_L}\nu_R \tilde{\Phi}_2$, whose smallness might be associated to the small induced vev of the additional $SU(2)$ scalar doublet as in the type-II seesaw for Dirac neutrinos~\cite{Bonilla:2016zef}.
In order to account for all neutrino oscillation observables the model can be further extended with extra RHNs neutral or not under $U(1)_{R}$. The actual construction of the neutrino sector is out of the scope of this paper.

\subsection{New interactions}

The relevant operators and scalar potential are listed next. The fermion interactions with
the $Z'$ are given by

\begin{eqnarray}
\mathcal{L}_{\text{gauge}}&\supset& 
(g'q_{\nu})Z'^{\mu}\left[
\overline{u_{R}}~\gamma_{\mu}u_{R}
-\overline{d_{R}}~\gamma_{\mu}d_{R}
-\overline{\mu_{R}}~\gamma_{\mu}\mu_{R}
+\overline{\nu_{R}}~\gamma_{\mu}\nu_{R}\right]~.\label{eq:muonGauge}
\end{eqnarray}

Then, the overall right- and left-handed couplings $C_{R,L}$ to the muon become $C_{R}=-g'$ and $C_{L}=0$ for $q_{\nu}=1$. We normalize the vector ($V$) and axial vector ($A$) couplings as $C_{V,A}=\tfrac{1}{2}(\pm C_{L}+C_{R})$~. The gauge-invariant Yukawas of the SM fermions are
\begin{align}
-\mathcal{L}_{Y}
&= y_{ie} \overline{L}_i e_R \Phi_1 + y_{i\mu} \overline{L}_i \mu_R \Phi_2+ y_{i\tau} \overline{L}_i \tau_R \Phi_1 + y_{i\nu}  \overline{L}_i \nu_R \widetilde{\Phi}_2 \notag\\
&+ y_{id} \overline{Q}_i d_R \Phi_2 + y_{is} \overline{Q}_i s_R \Phi_1+ y_{ib} \overline{Q}_i b_R \Phi_1 \notag\\
&+ y_{iu} \overline{Q}_i u_R \widetilde{\Phi}_2 + y_{ic} \overline{Q}_i c_R \widetilde{\Phi}_1+ y_{it} \overline{Q}_i t_R \widetilde{\Phi}_1
  + \text{H.c}. \label{eq:LYukawa}
\end{align}
with $i=1,2,3$ denoting the fermion generations and where
\begin{equation}
\Phi_{a}=\left(\begin{array}{c} 
\varphi_a^{+} \\
\varphi_a^{0}
\end{array}\right)=\left(\begin{array}{c}
\varphi_a^{+} \\
\dfrac{v_{\Phi_{a}}+\varphi_{aR}^{0}+i\varphi_{aI}^{0}}{\sqrt{2}}
\end{array}\right)~,~~~~~\text{with }a=1,2~.
 \end{equation}

 Regarding the generation of the muon mass, we consider the Yukawa matrix for charged lepton as diagonal, so that, $y_{2\mu}\neq0$.
 Thus from (\ref{eq:LYukawa}) $m_{\mu}=y_{2\mu}v_{\Phi_{2}}/\sqrt{2}$. A \textit{natural} muon Yukawa occurs if one selects $v_{\Phi_{2}}\approx m_{\mu}$ within one order of magnitude. Then $0.01\lesssim v_{\Phi_{2}}/\text{GeV}\lesssim1$ is equivalent to $0.1\lesssim y_{2\mu}\lesssim10$, with the possibility of saturating $y_{2\mu}=4\pi$. The sizability of $v_{\Phi_{2}}$ (or equivalently $y_{2\mu}$) will prove to be important in Sec. \ref{sec:gminus2} to accomodate the $(g-2)_{\mu}$ bounds. The scalar potential of our Two-Higgs doublet model (2HDM) plus complex singlet reads
\begin{align}
V(\Phi_{1},\Phi_{2},s)_{q_{s}=-1/2}
&= -\mu_{1}^{2}\Phi_{1}^{\dag}\Phi_{1} - \mu_{2}^{2}\Phi_{2}^{\dag}\Phi_{2} - \mu_{s}^{2}s^{*}s
+\lambda_{1}(\Phi_{1}^{\dag}\Phi_{1})^{2} + \lambda_{2}(\Phi_{2}^{\dag}\Phi_{2})^{2} + \lambda_{s}(s^{*}s)^{2} \notag \\
&+ \lambda_{12}\Phi_{1}^{\dag}\Phi_{1}\Phi_{2}^{\dag}\Phi_{2} + \lambda'_{12}\Phi_{1}^{\dag}\Phi_{2}\Phi_{2}^{\dag}\Phi_{1} + \lambda_{1s}\Phi_{1}^{\dag}\Phi_{1}s^{*}s + \lambda_{2s}\Phi_{2}^{\dag}\Phi_{2}s^{*}s+ \kappa (s^{2}\Phi_{1}^{\dag}\Phi_{2}+\text{h.c.})~,
\end{align}
where the $U(1)_R$ charge of the scalar singlet is taken as $q_s=-1/2$. For this charge choice, the allowed four-scalar operator with the Higgs doublets is $s^{2}\Phi_{1}^{\dag}\Phi_{2}$, with dimensionless coefficient $\kappa$ as in Ref.~\cite{Bonilla:2016zef}. An alternative, not pursued here, is choosing $q_{s}=-1$ under which the allowed four-scalar operator would be $s\Phi_{1}^{\dag}\Phi_{2}$ with a dimensionful mass parameter as coefficient.

\section{The charge radius discrepancy}
\label{sec:radius}

 In our model the contributions to the proton charge radius come from the exchange of the mediators $Z'$, $\Phi_{2}$ and $s$ between muons and protons. We later discuss why the scalar contributions are subleading/negligible. Furthermore, since left-handed chiral neutrinos do not interact with the nucleus through the $Z'$, they automatically avoid neutrino-nucleon scattering constraints \cite{Batell:2011qq}.

 The energy splitting in muonic hydrogen by exchange of the vectorial ($V$) component of a spin-1 mediator is given by \cite{Carlson:2012pc}
\begin{equation}
|\Delta E^{(V)}| = \dfrac{\bigl|~C_{V}^{\mu}C_{V}^{p}~\bigr|}{4\pi}\dfrac{m_{V}^{2}(m_{\text{red}}\alpha_{\text{em}})^{3}}{2(m_{V}+m_{\text{red}}\alpha_{\text{em}})^{4}}~, \label{eq:r2pEnergyV}
\end{equation}
with the analog $\Delta E^{(S)}$ by CP-even scalar exchange\footnote{Pseudoscalar contributions are tiny according to \cite{Carlson:2012pc}.} sharing the same expression. Above, the $C_{V}^{\mu(p)}$ are the vectorial muon (proton) couplings to the gauge boson, $m_{\text{red}}\equiv m_{p}m_{\mu}/(m_{\mu}+m_{p})$ is the muon-proton reduced mass, $m_{V}$ is the mediator mass, and $\alpha_{\text{em}}$ is the QED constant. The Lamb energy shift between levels $2S_{1/2}$ and $2P_{1/2}$ depends on the radius $r$ (in fm) through the following relation \cite{Carlson:2015jba},
\begin{equation}
\Delta E(r)=206.0336-5.2275r^{2}+0.0332\text{MeV}~. \label{ref:EtoRadius}
\end{equation}
So we are using Eq.~(\ref{ref:EtoRadius}) to fit the proton charge radius discrepancy shown in Eq. (\ref{eq:rpeANDmu}). This will correspond to an energy shift of $\Delta E\approx 306.4~\mu\text{eV}$.

\begin{figure}[h!]
\centering
\includegraphics[scale=0.5]{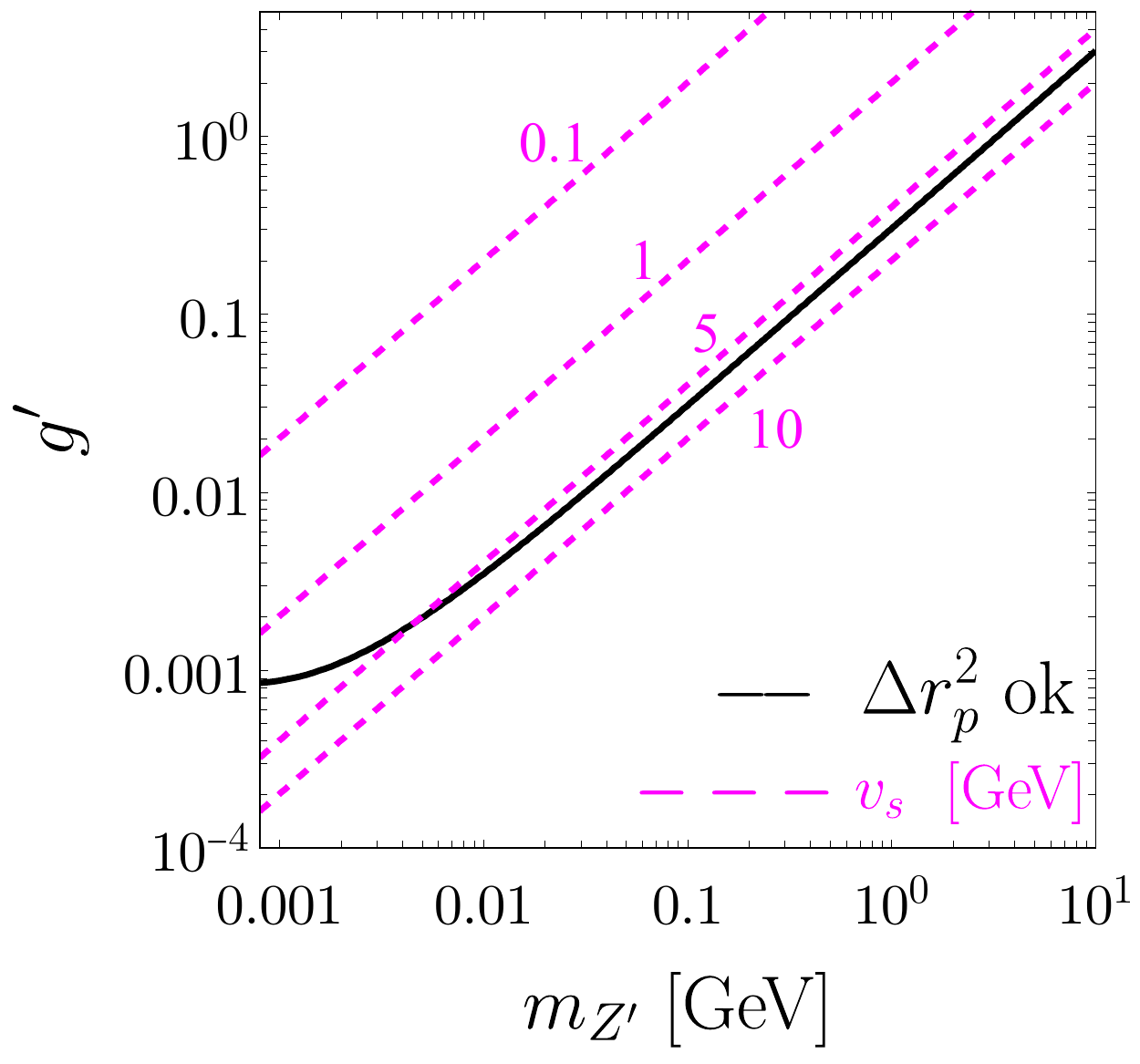}
\caption{$Z'$ mass-coupling curve fitting the $\Delta r_{p}^{2}$ discrepancy (solid black) obtained from equating the RHS of Eq. (6) to $306.4~\mu\text{eV}$, and the corresponding $U(1)_{R}$-breaking singlet vev (magenta dashed).}
\label{fig:gpMZvs}
\end{figure}
For purposes of simplicity, we take the $Z'$ coupling to the $u$ and $d$ quarks as an estimate of $C_{V}^{p}$, the vectorial $Z'$ coupling to the proton\footnote{The proportionality factor between quark-$Z'$ couplings and proton-$Z'$ couplings is determined through the sum of the constituent quark charges (vectorial) and through matrix elements techniques (axial).
}. Then from Eq.~(\ref{eq:muonGauge}), with $C_{V}^{\mu}=C_{V}^{p}\equiv C_{V}[Z']$, the relevant $Z'$ coupling appearing in the energy shift (\ref{eq:r2pEnergyV}) is $|C_{V}[Z']|=g'/2$. In addition, in (\ref{eq:r2pEnergyV}) $m_{V}=m_{Z'}$, which in our setup is $m_{Z'}\approx g'\sqrt{(v_{s}^{2}/4)+v_{\Phi_{2}}^{2}}$~.

The square of this value is the diagonal entry of the new gauge boson in the full $3\times3$ interaction-basis mass matrix of the neutral vector bosons after the Weinberg rotation. In the limit of vanishing kinetic mixing, this value is a good approximation for $m_{Z'}$ because the off-diagonal mass mixing of the $U(1)_{Y}$ and $U(1)_{R}$ gauge bosons goes as $-\tfrac{1}{2}g'\sqrt{g_{1}^{2}+g_{2}^{2}}v_{\Phi_{2}}^{2}$ and this is much smaller than $|m_{Z'}^{2}-m_{Z}^{2}|$ (see for example the paramametrization of Ref. \cite{Lindner:2018kjo}).

On the $(m{_Z'},g')$ plane in Fig. \ref{fig:gpMZvs}, a solid black curve satisfying $|\Delta E^{(V)}|\approx 306.4~\mu\text{eV}$ is shown, together with a few contours of the corresponding $v_{s}$ singlet vev (magenta dashed), which we will assume dominates the $Z'$ mass, i.e $m{_Z'}\approx g'v_{s}/2 $. These dashed contours indicate that $5\lesssim v_{s}/\text{GeV}\lesssim10$ works well for $\Delta r_{p}^{2}$, except when $m_{Z'}$ gets as low as 1 MeV\footnote{This curve's behavior is due to the constant term in denominator of Eq.~(\ref{eq:r2pEnergyV}).}.

 With the previous information it is suggested that the proton radius puzzle can be addressed in this model thanks $only$ to existence of a light $Z'$ that interacts only with one family of RH-quarks and -leptons. However, the assumption on the vevs ($v_s\sim\text{few GeVs}\gg v_{\Phi_2}\sim m_\mu$) has an impact on the scalar mass spectrum which is constrained by current experimental data. Moreover, given the features of the model, we will also consider the latest muon $(g-2)$ results, which provide important parameter space restrictions and predictions.

\section{Model constraints and results}
\label{sec:gminus2}

In what follows we will describe the experimental limits applied to our setup.
As discussed earlier, from Fig. \ref{fig:gpMZvs}, and under our choice $q_{s}=-1/2$, a vev $v_{s}\approx 6.5\text{ GeV}$ fits $\Delta r_{p}^{2}$ for $m_{Z'}$ above 1 MeV. This and other parameters choices will be collected in Table \ref{tab:bench}, where subindices will be dropped for the muon Yukawa, $y\equiv y_{2\mu}$, to ease notation. Now we proceed to discuss the scalar sector and its experimental restrictions.

\subsection{Limits on the scalar mass spectrum}
\label{sec:results}

The singly-charged Higgs from the $\Phi_{2}$ doublet is subject to severe LHC constraints. In the context of a muonphilic 2HDM, a bound $m_{H^{+}}\gtrsim 640\text{ GeV}$ has been reported \cite{Abe:2017jqo}. We choose two representative benchmarks, one with the $m_{H_{3}},m_{A},m_{H^{+}}$ sitting right at this bound, and another, more conservative one where they sit at 1 TeV. In order to have a grip on the $s$-like and $\varphi_{2}$-like scalar masses, we look closer at the mass matrices in a regime of small mixing. For a $O(1)$, sizable $\kappa$ coupling and under the vev hierarchy $v_{\Phi_{1}}\gg v_{s}\gg v_{\Phi_{2}}$, the $M_{R}^{2}$ mixing matrix in (\ref{eq:M2Rfull}) can always be brought to an approximate block-diagonal form
\begin{equation}
M_{R}^{2}\sim \left(\begin{array}{ccc}
2\lambda_{1}v_{\Phi_{1}}^{2} & \varepsilon           & \varepsilon' \\
\varepsilon           & 2\lambda_{s}v_{s}^{2} & -\kappa v_{\Phi_{1}}v_{s} \\
\varepsilon'          & -\kappa v_{\Phi_{1}}v_{s}    & \kappa v_{\Phi_{1}}v_{s}^{2}/(2v_{\Phi_{2}})
\end{array}\right) \label{eq:approxM2R}
\end{equation}
under a suitable choice of quartics. Above, $\varepsilon,\varepsilon'$ entries are no larger than the $(1,1)$ entry, but much smaller than the $(3,3)$ and $(2,3)$ entries. Then the diagonalization of the lower-right $2\times2$ block in $(\sigma_{R},\varphi_{2R}^{0})$ approximates the light $s$-like and $\varphi_{2}$-like masses, and it does it in terms of $\lambda_{s}~,\kappa~,v_{\Phi_{2}}$ only (recall that $v_{\Phi_{1}}^{2}=(246\text{ GeV})^{2}-v_{\Phi_{2}}^{2}$, and that $v_{s}$ is fixed by $\Delta r_{p}^{2}$). The approximate eigenvalues of this block are
\begin{equation}
m_{\text{light},\text{heavy}}^{2}=\tfrac{1}{2}\left( 2\lambda_{s}v_{s}^{2} + \kappa v_{\Phi_{1}}v_{s}^{2}/(2v_{\Phi_{2}}) \right) \mp \sqrt{ \tfrac{1}{4}\left[ 2\lambda_{s}v_{s}^{2}-\kappa v_{\Phi_{1}}v_{s}^{2}/(2v_{\Phi_{2}}) \right]^{2}+(-\kappa v_{\Phi_{1}}v_{s})^{2} }~.
\end{equation}
At next-to-leading order in $v_{\Phi_{2}}$, the light mass is 
\begin{equation}
m_{\text{light}}\approx \sqrt{2(\lambda_{s}v_{s}^{2}-\kappa v_{\Phi_{1}}v_{\Phi_{2}})}~,
\end{equation}
whereas the heavy one is well approximated by the diagonal entry in $\varphi_{2R}$ at leading order in $v_{\Phi_{2}}$
\begin{equation}
m_{\text{heavy}}\approx \sqrt{ \dfrac{\kappa}{2}\dfrac{v_{\Phi_{1}}}{v_{\Phi_{2}}} }v_{s}~. \label{eq:heavier}
\end{equation}
The charged scalar mass is also approximated by (\ref{eq:heavier}), and this is also true for the pseudoscalar $A$ too. Therefore, the $H_{3},A,H^{+}$ in this regime are highly degenerate. Given these masses, $m_{\text{heavy}}$ is freed from the 640 GeV bound by LHC on charged scalars if
\begin{equation}
\sqrt{\kappa/v_{\Phi_{2}}}\gtrsim 8.88\text{ GeV}~. \label{eq:LHCCond}
\end{equation}
Regarding the $s$-like singlet scalar, its mass $m_{\text{light}}$ must be rather light in order to help the total $\Delta a_{\mu}$ reach the current limit. Since its coupling to muons is supressed by the small $s$-$\Phi_{2}$ mixing, $y$ must be large enough to compensate. At light mass values, however, a tachyonic $m_{\text{light}}$ must be avoided. This is achieved if
\begin{equation}
\bigl| \lambda_{s}v_{s}^{2}-\kappa v_{\Phi_{1}}v_{\Phi_{2}} \bigr|\geq0~.\label{eq:tachyonCond}
\end{equation}

\begin{figure}[h!]
\centering
\includegraphics[scale=0.45]{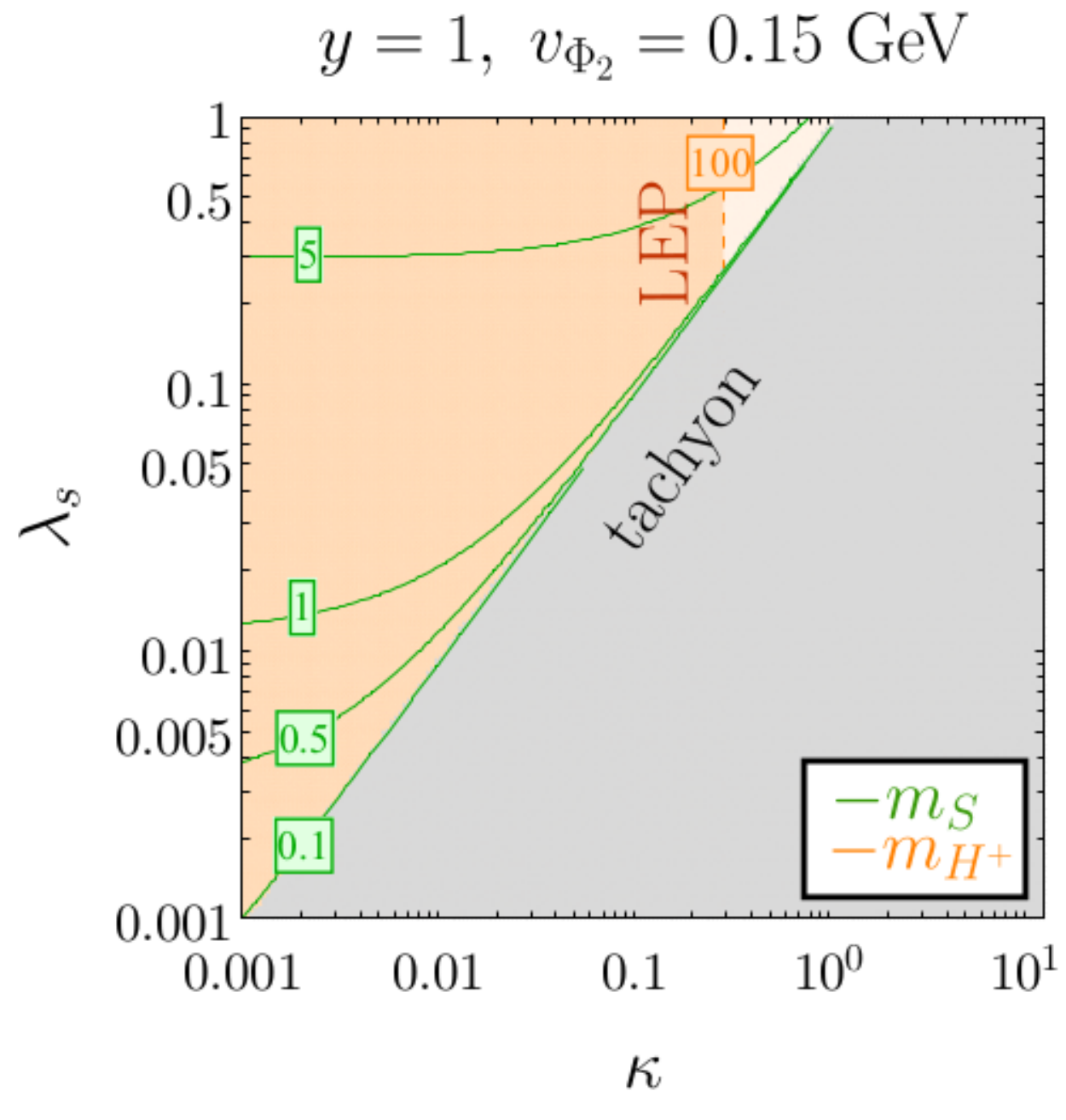}
\hspace{5mm}
\includegraphics[scale=0.45]{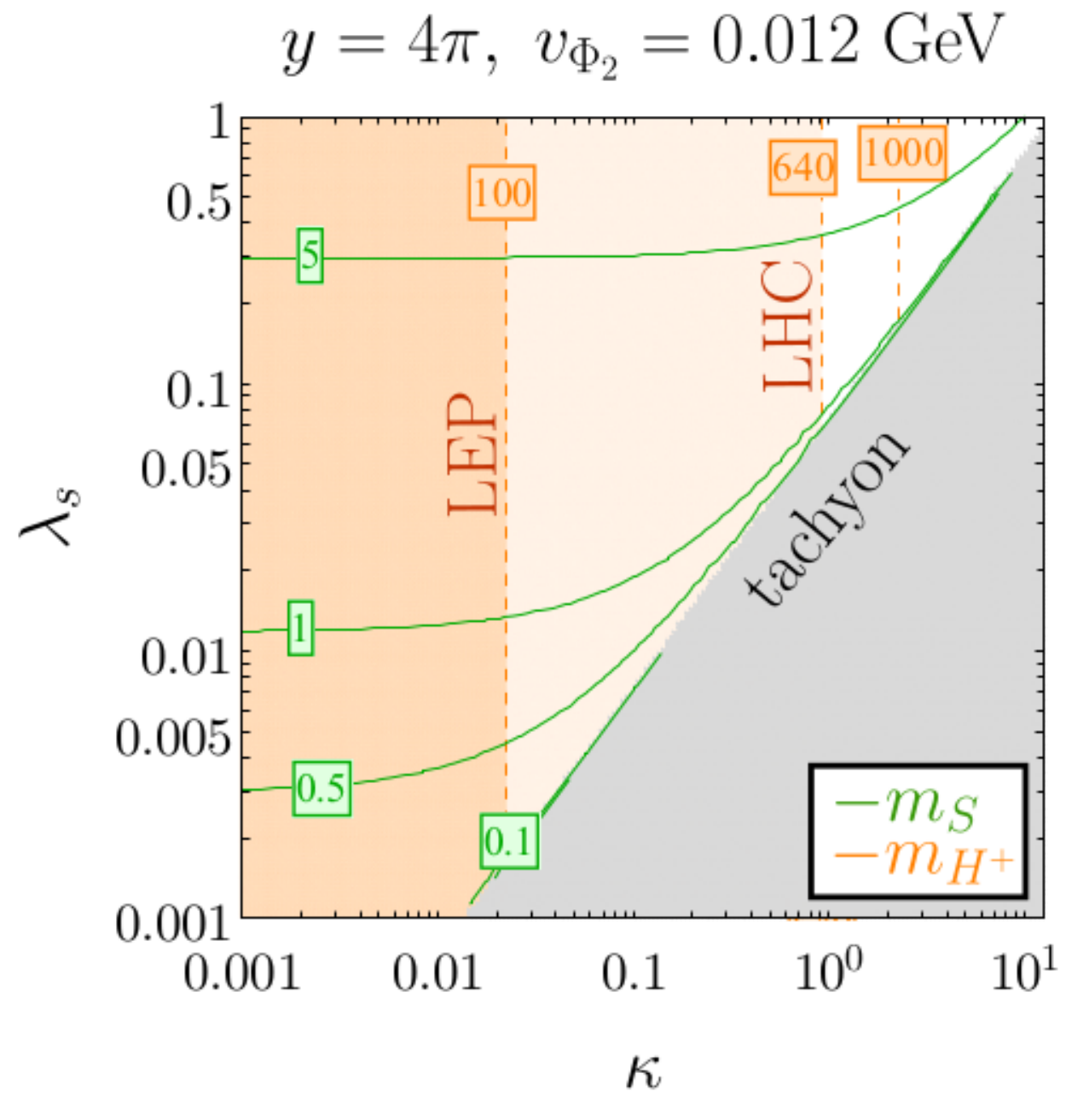}
\caption{Contours of $s$-like (green) and $\varphi_{2}$-like (orange) scalar masses at two $v_{\Phi_{2}}$ choices (i.e. at two $y$ muon Yukawa choices). The LHC limit on muonphuilic $H^{+}$ scalars is shown, which pushes $y$ to large values near the perturbative $4\pi$ value.}
\label{fig:kappaLas}
\end{figure}

Contours of the light singlet (solid green) and charged scalar (dashed orange) masses are shown in the panels of Fig. \ref{fig:kappaLas} in the $(\kappa,\lambda_{s})$ plane, at two representative, fixed Yukawa values (i.e. fixed $v_{\Phi_{2}}$), $y=$1 and $y=4\pi$. A tachyonic $m_{\text{light}}$ for the singlet is developed over the gray region. A noticeable feature is that the new doublet vev must be sufficiently smaller than $v_{\Phi_{2}}\sim0.1\text{ GeV}$ for the singlet mass contours to be freed from the LHC limit on muonphilic $H^{+}$, which pushes $y$ to values larger than 1. For reference, the conservative LEP constraint on $H^{+}$ is shown as well. 

\begin{figure}[h!]
\centering
\includegraphics[scale=0.45]{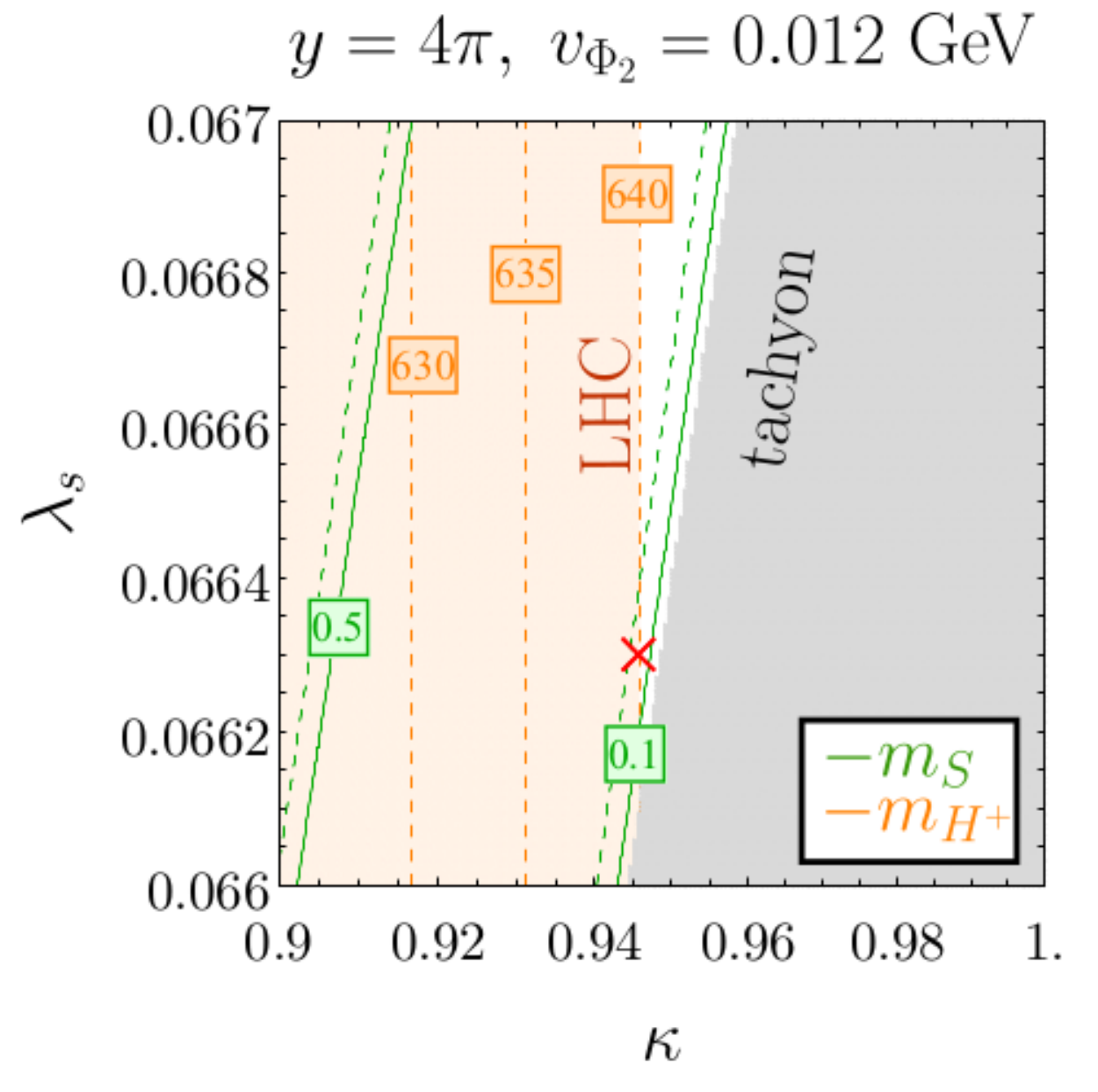}
\hspace{5mm}
\includegraphics[scale=0.45]{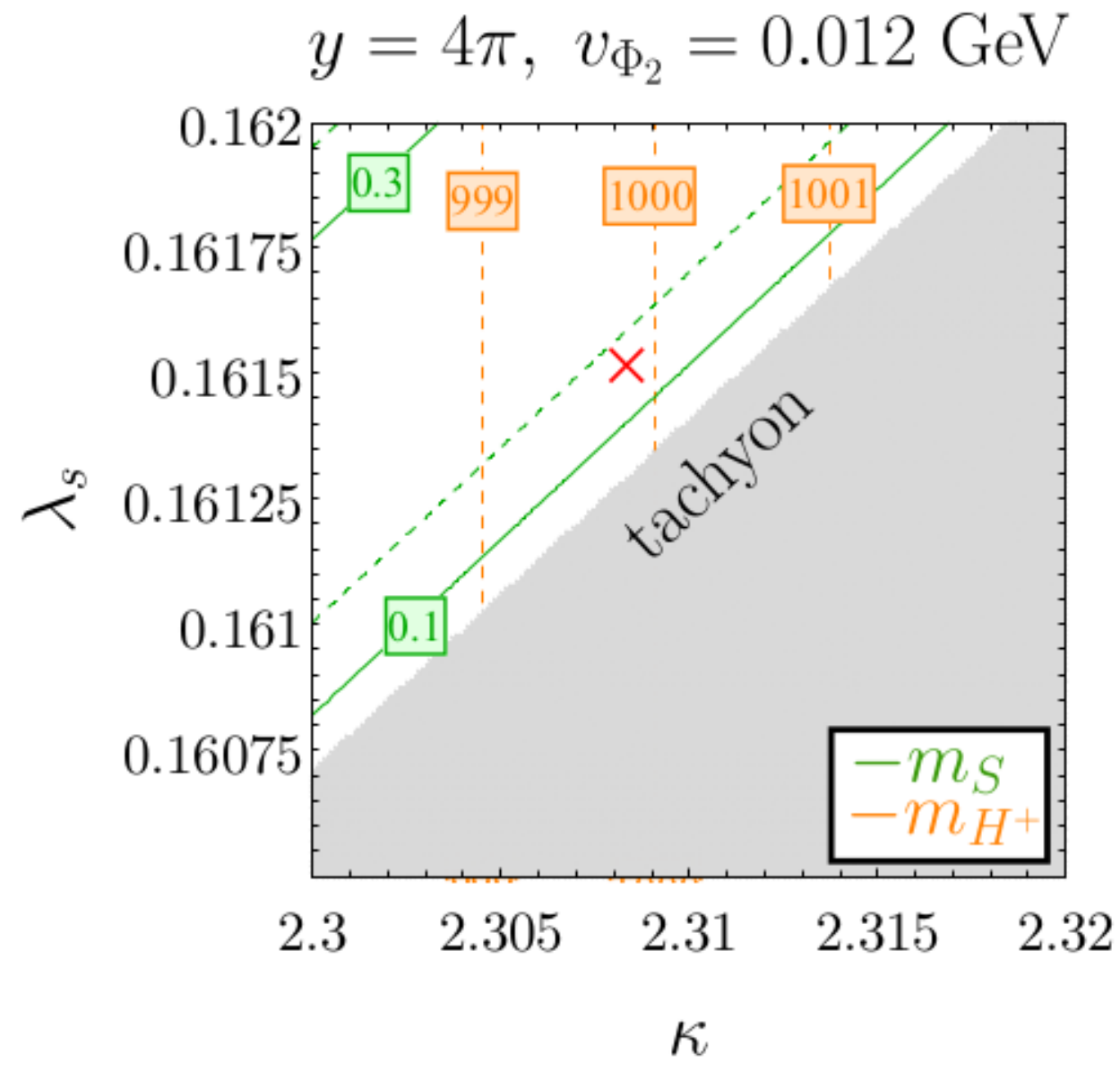}
\caption{Zooms of Fig. \ref{fig:kappaLas} right panel, one around the LHC limit on $H^{+}$ scalars (left) and the other around the $m_{H^{+}}=1\text{ TeV}$ mass contour. Red crosses mark the benchmarks in Table \ref{tab:bench}.}
\label{fig:kappaLasZOOM}
\end{figure}

Two zooms of the right panel of Fig. \ref{fig:kappaLas} are displayed in Fig. \ref{fig:kappaLasZOOM}. The dashed green contours in the panels are $m_{\text{light}}$ values from the full $3\times3$ mixing matrix. The deviation with respect to the contours of the $2\times2$ subblock (solid green) is small as long as the approximate form (\ref{eq:approxM2R}) holds. One observes that regions with (non-tachyonic) sub-GeV $m_{\text{light}}$ and $m_{\text{heavy}}\gtrsim640\text{ GeV}$ are narrow and very sensitive to $\kappa$, $\lambda_{s}$. The model is then said to demand a singlet-like scalar in a well-defined range of tens of MeV.

\begin{table}[h!]
\begin{tabular}{|cccc|}
\hline
$\lambda_{1}=0.13$     & $\lambda_{2}=0.13$  & $v_{s}=6.5\text{ GeV}$           & \\
$\lambda_{1s}=0.01$    & $\lambda_{2s}=0.01$ & $\lambda_{12}=\lambda_{12}'=0.1$ & \\
\hline
Benchmark \textbf{A}                    & $\kappa=0.4677$ & $\lambda_{s}=0.0329$ & $y=4\pi$ \\
& $(m_{H_{1}},~m_{H_{2}},~m_{H_{3}})\approx(75\text{ MeV},~125\text{ GeV},~640\text{ GeV})~$ & & \\
\hline
Benchmark \textbf{B} & $\kappa=2.308$   & $\lambda_{s}=0.1615$ & $y=4\pi$ \\
& $(m_{H_{1}},~m_{H_{2}},~m_{H_{3}})\approx(80\text{ MeV},125\text{ GeV},1\text{ TeV})~$ & & \\
\hline 
\end{tabular}
\caption{Scalar sector benchmarks. Parameters in the top cell are shared by both benchmarks.}
\label{tab:bench}
\end{table}

We list the mass spectrum at two benchmarks in Table~\ref{tab:bench}, and mark them with red crosses in the panels of Fig.~\ref{fig:kappaLas}. Notice that the charge scalar mass $m_{H^{+}}$ sits at either 640 GeV or 1 TeV, with $m_{A}\approx m_{H^{+}}\approx m_{H_{3}}$ in both benchmarks. We follow the mass ordering $m_{H_{1}}\leq m_{H_{2}}\leq m_{H_{3}}$, hence $H_{1}$ is a sub-GeV $s$-like light state, $H_2$ the SM-like Higgs, and $H_3$ is a heavier scalar (with the physical states mostly unmixed). 

Having established the scalar mass spectrum one can also determine the size of the 
scalar-fermion couplings, such as, the scalar-muon and -quark couplings which are needed to compute the contribution of the scalars to the charge radius $\Delta r_{p}^{2}$. From Eq.(\ref{eq:LYukawa}), the scalar-muon couplings are given by
\begin{equation}
C^{\mu}_{S}[H_{k}]=-\dfrac{y}{\sqrt{2}}(\mathcal{O}_{R})_{k3}~\ \ \text{with}\ \ k=1,2,3,
\label{eq:couplingSmu}
\end{equation}
where $\mathcal{O}_{R}$ is the CP-even mixing matrix and is formally determined from Appendix~\ref{sec:appA}. 
Meanwhile, the scalar couplings to the proton are given by
\begin{equation}
C_{S}^{p}[H_{k}]=-\dfrac{y_{q}}{\sqrt{2}}(\mathcal{O}_{R})_{k3} \ \ \text{with}\ \ k=1,2,3~, \label{eq:couplingSq}
\end{equation}
where $y_{q}$ is the quark Yukawa. Using Eqs.~(\ref{eq:couplingSmu}) and~(\ref{eq:couplingSq}), one can write one coupling in terms of the other as $C_{S}^{q}= \left(y_{q}/y_{} \right) C_{S}^{\mu} \sim \left(m_u/m_\mu \right)C_{S}^{\mu} \approx 2\times10^{-2} C_{S}^{\mu}$. Numerically, at the benchmarks
\begin{equation}
\bigl( C^{\mu}_{S}[H_{1}],~C^{\mu}_{S}[H_{2}],C^{\mu}_{S}[H_{3}] \bigr)\approx (-0.03,~-4.8\times10^{-4},~-8.9)~,~~~~~C^{\mu}_{P}[A]\approx \widetilde{C}^{\mu}_{S,P}[H^{+}]\approx C^{\mu}_{S}[H_{3}]~\ \ \text{in both \textbf{A} and \textbf{B}} \label{eq:mixings}
\end{equation}
The value of these couplings stays almost identical in either benchmark, since $y=4\pi$ in both.\\

Plugging the resulting couplings and scalar masses into $\Delta E^{(S)}$, as in Eq.~(\ref{eq:r2pEnergyV}), one gets that the Higgs-like  $H_2$ and $H_3$ provide a negligible contribution to $\Delta r_{p}^{2}$. The sub-GeV scalar $H_1$ adds to the required energy shift at the level of 10\%, representing a
subleading contribution to the charge radius. This makes plausible to stick to the limit in which the $\Delta r_{p}^{2}$ deviation is addressed by $Z'$ alone\footnote{In contrast to our model, there exist scenarios that tackle $r_{p}$ and the muon $(g-2)$ without a UV completion. See for instance Ref. \cite{Liu:2016qwd} where a light electrophobic scalar fits $r_{p}$ and the muon $(g-2)$ while satisfying related bounds due to its coupling to neutrons.}.

\subsection{The muon anomalous magnetic moment}

As we have seen, our model is characterized by its mounphilic interactions, which give additional corrections to the muon anomalous magnetic moment, commonly parametrized as $\Delta a_{\mu}\equiv (g-2)_{\mu}/2$. Therefore, in what follows we show the allowed parameter space where our model simultaneously satisfies $\Delta r_{p}^{2}$, the limits on the new scalars, and the constraints coming from latest $(g-2)_{\mu}$ results.

In light of the latest $(g-2)_{\mu}$ Fermilab measurement \cite{Abi:2021gix,Albahri:2021kmg}, we interpret these results either as a bound on BSM states under the assumption that the SM deviation eventually fades away or as a hint of new physics. This is done by substracting the $(g-2)_{\mu}$ measurement to two theory predictions, namely
\begin{align}
\Delta a_{\mu}^{\text{latt}}/10^{-11} &= 109\pm71~, \label{eq:gMinus2nonpert} \\
\Delta a_{\mu}^{\text{SM,\text{th}}}/10^{-11} &= 251\pm59~. \label{eq:gMinus2pert}
\end{align}
$\Delta a^{\text{latt}}$ is the current Fermilab+BNL average after substracting the latest QCD theoretical prediction by lattice including vacuum hadron polarization \cite{Borsanyi:2020mff}. A summary of more lattice results by other groups is discussed in Refs. \cite{Gerardin:2020gpp,Davier:2019can}. On the other hand, $\Delta a_{\mu}^{\text{SM,\text{th}}}$ is the current Fermilab+BNL average after substracting the theoretical SM contribution \cite{Aoyama:2020ynm}.

\begin{figure}[h!]
\centering
\includegraphics[scale=0.45]{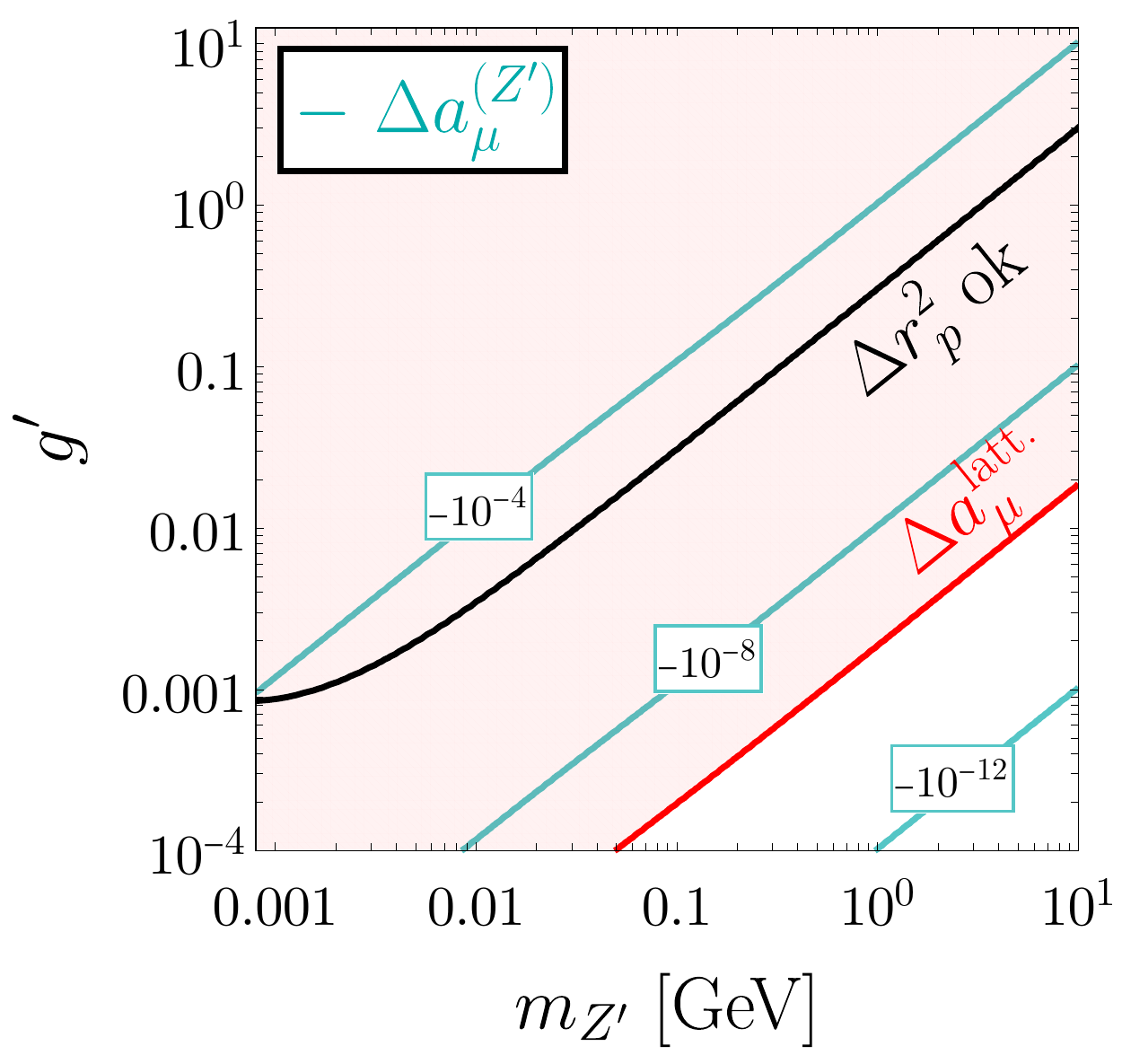}
\caption{
$(g-2)_{\mu}$ contribution by a $Z'$ coupled to RH fermions (cyan contours). The $\Delta r_{p}^{2}$ fitting curve is depicted in dashed black. The $(g-2)_{\mu}$ lattice bound exclusion from Eq. (17) is shaded red.
}
\label{fig:amu}
\end{figure}

In this scenario, $\Delta a_{\mu}$ receives contributions from all physical scalars and the new gauge boson, all of them are listed in Appendix~\ref{sec:appB}. Since the $Z'$ couples exclusively to RH fermions,
\begin{equation}
C_{V}[Z']=C_{A}[Z']=-g'/2, \label{eq:couplingsZp}
\end{equation}
the contribution to $(g-2)_{\mu}$ is negative, given that the axial piece of $\Delta a_{\mu}$ is approximately 10 times larger that the vectorial
one at equal couplings~\cite{Jegerlehner:2009ry}. Fig.~\ref{fig:amu} shows $\Delta a_{\mu}^{(Z')}$ for different $Z'$ masses and gauge couplings, displayed as solid blue contours in the $(g',m_{Z'})$-plane. According to the lattice bound (solid red) in Eq. (\ref{eq:gMinus2nonpert}),
$\Delta a_{\mu}^{(Z')}<0$ is not excluded as long as its magnitude lies within its $2\sigma$ range, this would correspond to $\Delta a_{\mu}$ contours outside the red shaded region.
However, notice that values allowed from the lattice constraint are incompatible with $\Delta r_{p}^{2}$ (solid black line). In other words, the safe region from the lattice constraint has smaller $r_{p}$ than the required one to fit the charge radius discrepancy, as noticed in Refs. \cite{Carlson:2012pc,Kirpichnikov:2020tcf} .

For the scalar contributions to $\Delta a_{\mu}$ one needs to extract their couplings to muons. The $H_{1,2,3}$ couplings are already listed in Eq. (\ref{eq:couplingSmu}) and those of the physical pseudoscalar and the charged scalar are
\begin{equation}
C^{\mu}_{P}[A]=-\dfrac{y_{}}{\sqrt{2}}(\mathcal{O}_{I})_{33}~,~~~~~\widetilde{C}^{\mu}_{S,P}[H^{+}]=-\dfrac{1}{2}(-y_{}\pm y_{\nu})(\mathcal{O}_{\pm})_{22}~, \label{eq:couplingsPandTilde}
\end{equation}
expressed in terms of the corresponding mixing matrices given in Appendix~\ref{sec:appA}. Since $y_\nu\sim m_\nu/v_{\Phi_2}\sim10^{-8}\ll y$, the couplings $\widetilde{C}^{\mu}_{S}[H^{+}]$ and $\widetilde{C}^{\mu}_{S}[H^{+}]$ are practically the same which leads to a negligible contribution from the charged scalar to $\Delta a_{\mu}$.

As we have seen, the parameter $\kappa$ controls the masses of the heavy scalars and tends to be of order $\mathcal{O}(1)$, look at Fig.~{\ref{fig:kappaLas}} for reference. This leads to a high degree of degeneracy of the $\varphi_{2}$-like scalars. Then, the contributions to $(g-2)_{\mu}$ by near-degenerate CP-even and CP-odd  scalars partially cancel among each other, for comparable couplings. This is similar to what occurs with the contribution from the charged scalar \cite{Jegerlehner:2009ry}.
As a result, the only sizable scalar contribution to the anomalous magnetic moment of the muon comes from the $s$-like state $H_1$ with sub-GeV mass. We then look for regions of the parameter space  where its contribution to $(g-2)_{\mu}$ together with that of the $Z'$ satisfies the experimental constraints
as well as explain the charge radius.\\

\begin{figure}[h!]
\centering
\includegraphics[scale=0.5]{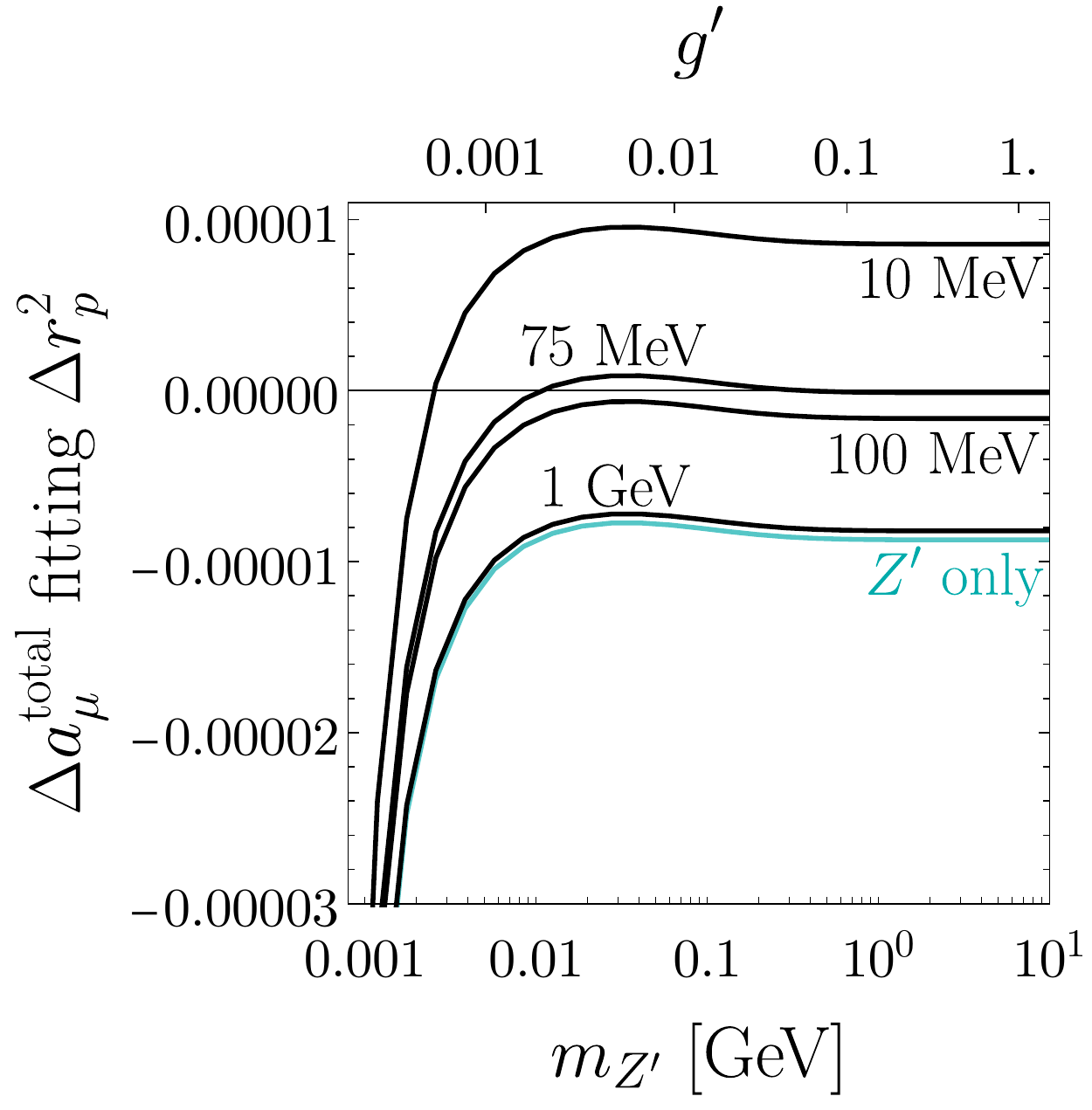}
\hspace{5mm}
\includegraphics[scale=0.5]{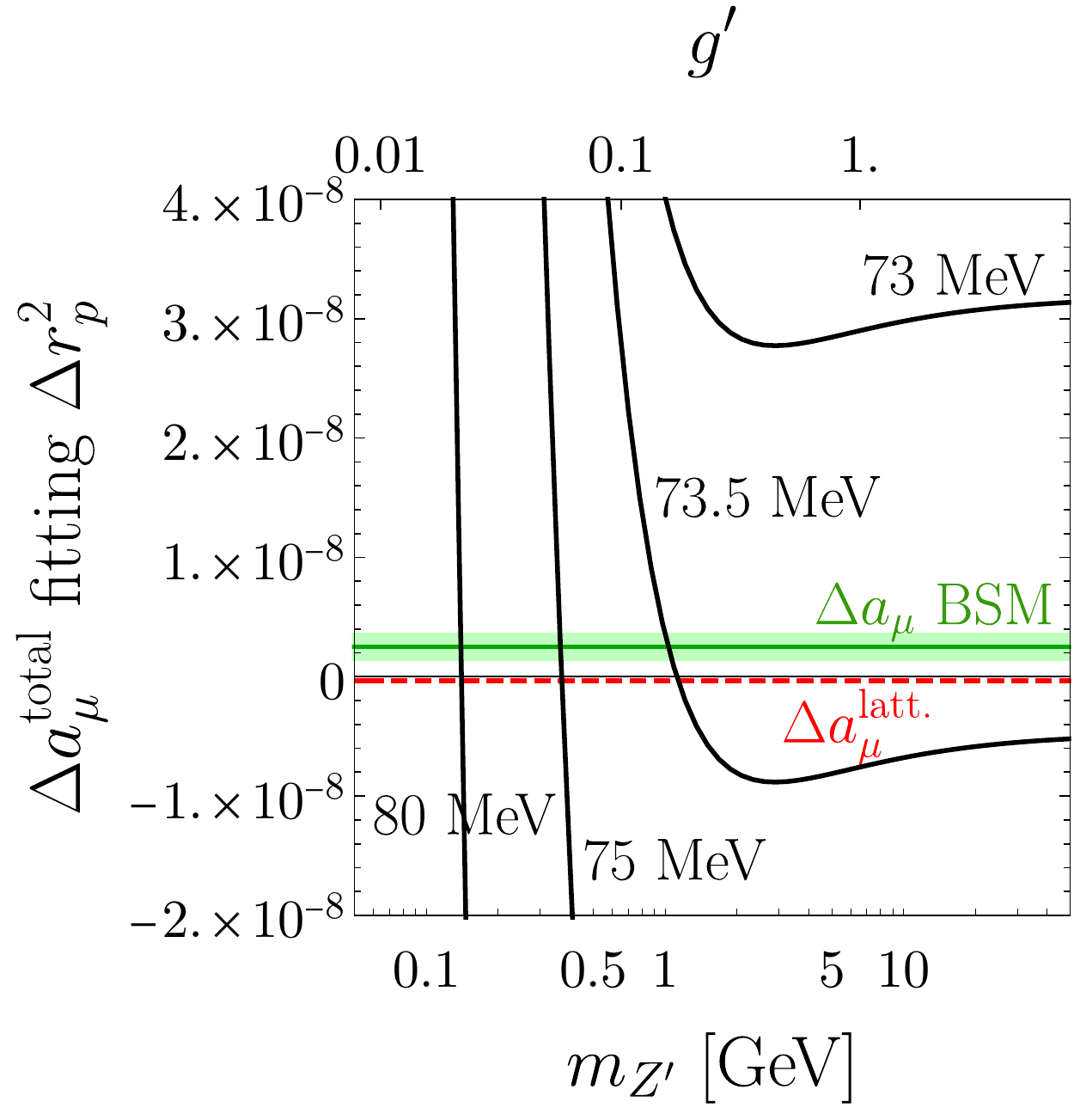}
\caption{\textbf{Left:} $Z'$-only (blue curve) and joint $Z'+s$ contribution to $(g-2)_{\mu}$ (black curves). The vertical $g'$ ticks indicate corresponding gauge coupling values fitting $\Delta r_{p}^{2}$ at each $m_{Z'}$. \textbf{Right:} Zoomed region for the $(g-2)_{\mu}$ lattice bound (dashed red) and $2\sigma$ band of the new physics hint (green).}
\label{fig:gMinus2Master}
\end{figure}

The left panel of Fig.~\ref{fig:gMinus2Master} plots $\Delta a_{\mu}$ by the $Z'$ alone (in blue) as a function of its mass, where at each $m_{Z'}$ the $g'$ coupling shown in the top axis fits $\Delta r_{p}^{2}$ according to Fig. \ref{fig:gpMZvs}. Even though $g'$ is varied, the required $v_{s}$ for the corresponding $m_{Z'}$ in the bottom axis stays constant at the benchmark value as long as $m_{Z'}$ is above a few MeV. From the graph it is clear that $\Delta a_{\mu}^{(Z')}$ alone is too large and negative to hit either the $(g-2)_{\mu}$ lattice bound on new physics or the $2\sigma$ band of the tentative signal. Here is where the light $s$-like scalar remedies this situation: the black curves in the same panel indicate the combined $\Delta a_{\mu}$ by $Z'$ and $s$, evaluated at the scalar-muon couplings (\ref{eq:mixings}). One sees that for tens of MeV it is possible for the black curves to hit the region relevant for $(g-2)_{\mu}$ bounds. The right panel of Fig. \ref{fig:gMinus2Master} zooms in into the lattice bound (red dashed) and signal hint (green band) ballpark. Therefore, the model is capable of simultaneously fit $\Delta r_{p}^{2}$ and be compatible with $(g-2)_{\mu}$ measurements, while generating the muon mass and evading current collider constraints on heavy scalars.

\subsection{Further restrictions on $Z'$}

Before concluding, we briefly mention constraints that may apply to our specific kind of $Z'$. First, our $Z'$ expects to face looser constraints compared to other types of $Z'$s that couple to electrons at tree-level\footnote{Coupling to $e^{+}e^{-}$ here is induced by kinetic mixing only.} . For instance, this $Z'$ automatically avoids the stringent limits from rare pion decays $\pi^{0}\to \gamma Z'\to \gamma(e^{+}e^{-})$ and $\nu-e$ scattering.

$\mu$-flavored $Z'$s are however subject to a variety of limits \cite{Bauer:2020itv}. For $m_{Z'}$ exceeding the $\mu^{+}\mu^{-}$ threshold, there are CMS and BaBar searches on $Z'Z'\to 4\mu$ as well as $B^{+}\to K^{+}Z'\to K^{+}(\mu^{+}\mu^{-})$ in LHCb. Below the $\mu^{+}\mu^{-}$ threshold there are $\pi^{0}\to \gamma Z'\to \gamma(\nu \nu)$ decay constraints. Due to possible kinetic mixing and its modification to $m_{Z'}$, there are also bounds from atomic parity violation and the electroweak $T$-parameter.

More importantly, there are current stringent constraints on the $(m_{Z'},\varepsilon)$ plane, where $\varepsilon=eg'/(16\pi^{2})$ is the kinetic mixing parameter. See for instance the limits by the NA64 and other collaborations in Ref. \cite{Banerjee:2019pds}. A simple analysis shows that the model is consistent in a wide region of the allowed parameter space. In addition, there are neutrino-related bounds from non-standard induced $\nu$-interactions, or coherent elastic neutrino nucleus scattering. At masses near $10\text{ MeV}$ there are also $\Delta N_{\text{eff}}$ constraints on dark radiation degrees of freedom. 

It is worth mentioning that a variety of nuclear experiments impose bounds on the ratio $C_{V}^{n}/C_{V}^{p}$ of neutron-$Z'$ to proton-$Z'$ coupling, where $C_{V}^{n}=C_{V}^{u}+2C_{V}^{d}$ and $C_{V}^{p}=2C_{V}^{u}+C_{V}^{d}$. For example, measurements on another muon-nucleus bound state, the Muonic Deuterium ($\mu\text{D}$), put important exclusion limits on this ratio. From Ref.~\cite{Liu:2016qwd}, one can see that results on $\mu\text{D}$ allow the window $-1\lesssim C_{V}^{n}/C_{V}^{p}\lesssim 0$ for $m_{Z'}<100\text{ MeV}$. In our setup, see Sec. \ref{sec:content}, given that the $U(1)_{R}$ charge assignments for $u$ and $d$ are fully determined by $q_{\nu}=1$, the ratio of neutron- and proton-$Z'$ couplings satisfies $C_{V}^{n}/C_{V}^{p}=-1$ for $m_{Z'}$ in the 10-100 MeV range. This small tension with the $\mu\text{D}$ constraint can be easily diluted if the $u$ and $d$ are placed in distinct anomaly-free sets. That is, for instance, taking another anomaly free solution $\{ u_{R},b_{R},\mu_{R},\nu_{R}\}$ with $U(1)_R$ charge parametrized by $q_{\nu}$ and $\{ c_{R},d_{R},\tau_{R},\nu'_{R}\}$ parametrized by $q'_{\nu}\neq q_{\nu}$, then $C_{V}^{n}/C_{V}^{p}=(2q_{\nu}-q'_{\nu})/(-2q'_{\nu}+q_{\nu})$ can be varied and brought to safety from the $\mu\text{D}$ limits while leaving our general conclusions for $\Delta r_{p}^{2}$ and $(g-2)_{\mu}$ unaffected \footnote{Certainly one more doublet $\Phi_{3}$ is required to provide masses to $c,d,\tau$, which simply duplicates the spectrum of heavy Higgses $H^{0},A^{0},H^{\pm}$. }. A detailed analysis of the multitude of bounds on $Z'$ is left for an upcoming work \cite{Alvarado:2021xyz}. The general $Z'$ constraints mentioned above must be implemented by taking into account 1) that $Z'$ is coupled to RH fermions only, and 2) that we have at our disposal some freedom in the quark sector to lock the RH quark rotations as diagonal matrices, and to turn off some $u$ and $d$ Yukawas to avoid dangerous $H^{+}$-mediated meson processes.

\section{Conclusions}
\label{sec:conclusion}

Every few years during the last decade muons have receive renewed attention from the community. This has been a consequence of the various experimental results that have hinted at new physics, among which two good examples are the $B$-meson decay anomalies and the muon $(g-2)$ measurement. Other less known instances, such as the proton charge radius, are also exciting venues for searches of discoveries beyond the Standard Model.

Before the seemingly surviving discrepancy in the proton charge radius succumbs to the improvement in precision techniques or better yet, a discovery is made, models accommodating it can be implemented and tested. This work proposes a SM extension based on a $U(1)_{R}$ gauge boson with exclusive coupling to right-handed muons, up and down quarks at tree-level, that addresses the $\Delta r_{p}$ discrepancy for sub-GeV $Z'$. Spontaneous breaking of the $U(1)_{R}$, the generation of the $\mu, u$, and $d$ masses, and agreement with the recent $(g-2)$ measurements demand extra scalar degrees of freedom. The new scalar states can be brought to consistency with LHC limits albeit in a tight region of the parameter space, with a light singlet-like scalar at the tens of MeV, and doublet-like, near-degenerate neutral, charged and pseudoscalar states no lighter than $\sim 640\text{ GeV}$. Further constraints arise for the gauge boson that deserve a study on its own, focusing on the preferential coupling of the $Z'$ to right-handed chiral fermions. 

\black
\begin{acknowledgments}
A.A. acknowledges support from SNI-CONACYT.
The work of C.B. has been supported by ANID under the FONDECYT grant ``Nu Physics'' No.  11201240. C.A. thanks the Facultad de Ciencias of the Universidad de Colima for its hospitality.
\end{acknowledgments}


\appendix

\section{Mass mixing and spectra}
\label{sec:appA}

\subsection{Scalar sector}

After acquiring vacuum expectation values (vevs) the fields are shifted as follows, 
 \begin{eqnarray}
  \varphi_i^0 = \frac{1}{\sqrt{2}}\left(v_{\Phi_i}+ \varphi_{R_1}+i \varphi_{I_1} \right), \ \
  \text{and}\ \
  s = \frac{1}{\sqrt{2}}\left(v_{s}+ s_R +i s_I \right)~,
 \end{eqnarray} 
 so the extremum conditions are
 \begin{eqnarray}\label{minconds}
  \mu_{1}^2&=&\frac{1}{2} \left(2 \lambda _1 v_{\Phi_1}^2+\lambda_{1s} v_s^2+\left(\lambda_{12}+\lambda'_{12}\right) v_{\Phi _2}^2-\frac{\kappa  v_s^2 v_{\Phi _2}}{v_{\Phi_1}}\right)~,\notag\\
  \mu_{2}^2&=&\frac{1}{2} \left(\left(\lambda_{12}+\lambda'_{12}\right) v_{\Phi_1}^2 +\lambda_{2s} v_s^2+2 \lambda _{2} v_{\Phi _2}^2-\frac{\kappa  v_{\Phi_1} v_s^2}{v_{\Phi _2}}\right)~,\\
  \mu_{s}^2&=&\frac{1}{2} \left(\lambda_{1s} v_{\Phi_1}^2 +\lambda_{2s} v_{\Phi _2}^2+2 \lambda_s v_s^2-2 \kappa  v_{\Phi_1} v_{\Phi _2}\right)~.\notag
  \end{eqnarray}
Upon minimization, the scalar mass matrices read
\begin{eqnarray}
 M_R^2=\left(
\begin{array}{ccc}
 2\lambda _1 v_{\Phi_1}^2 +\frac{\kappa  v_s^2 v_{\Phi _2}}{2 v_{\Phi_1}}      & \left(\lambda_{1s} v_{\Phi_1} -\kappa  v_{\Phi _2}\right) v_s       & \left(\lambda_{12}+\lambda'_{12}\right) v_{\Phi_1} v_{\Phi _2}  -\frac{\kappa  v_s^2}{2} \\
  \left(\lambda_{1s} v_{\Phi_1} -\kappa  v_{\Phi _2}\right)v_s   & 2 \lambda_s v_s^2                                            &  \left(\lambda_{2s} v_{\Phi _2} -\kappa  v_{\Phi_1}\right) v_s \\
  \left(\lambda_{12}+\lambda'_{12}\right) v_{\Phi_1} v_{\Phi _2}  -\frac{\kappa  v_s^2}{2}      & \left(\lambda_{2s} v_{\Phi _2} -\kappa  v_{\Phi_1}\right) v_s      & 2 \lambda _{2} v_{\Phi _2}^2 +\frac{\kappa  v_{\Phi_1} v_s^2}{2 v_{\Phi _2}} \\
\end{array} \label{eq:M2Rfull}
\right)
\end{eqnarray}
 and
 \begin{eqnarray}
\label{eq:imag}
 M_I^2=\kappa\left(
\begin{array}{ccc}
 \frac{v_s^2 v_{\Phi _2}}{2 v_{\Phi_1}} & -v_s v_{\Phi _2} & -\frac{v_s^2}{2} \\
 -v_s v_{\Phi _2} & 2 v_{\Phi_1} v_{\Phi _2} & v_{\Phi_1} v_s \\
 -\frac{v_s^2}{2} & v_{\Phi_1} v_s & \frac{v_{\Phi_1} v_s^2}{2 v_{\Phi _2}} \\
\end{array}
\right)~.
\end{eqnarray}

Define the rotation matrices $\mathcal{O}_{R}$ and $\mathcal{O}_{I}$ according to $\text{diag}(m_{H_{1}}^2,m_{H_{2}}^2,m_{H_{3}}^2)=\mathcal{O}_{R}M_{R}^2\mathcal{O}_{R}^T$
and $\text{diag}(0,0,m_{A}^2)=\mathcal{O}_{I}M_{I}^2\mathcal{O}_{I}^T$, then
\begin{equation}
m_A^2= \frac{\kappa  \left(v_{\Phi_1}^2 \left(v_{s}^2
+4 v_{\Phi_2}^2\right)+v_{s}^2 v_{\Phi_2}^2\right)}{2 v_{\Phi_1} v_{\Phi_2}}~,~~~~~
\mathcal{O}_{I} =\left(\begin{array}{ccc}
\alpha v_{\Phi_{1}} & 0 & \alpha v_{\Phi_2} \\
-2\alpha \beta v_{\Phi_{1}}v_{\Phi_{2}}^{2} & -\tfrac{\beta}{\alpha}v_{s} & 2\alpha \beta v_{\Phi_{1}}^{2}v_{\Phi_{2}} \\
-\beta v_{s}v_{\Phi_{2}} & 2\beta v_{\Phi_{1}}v_{\Phi_{2}} & \beta v_{\Phi_{1}}v_{s}
\end{array}\right)
\end{equation}
where
\begin{equation}
\alpha=\dfrac{1}{\sqrt{v_{\Phi_{1}}^{2}+v_{\Phi_{2}}^{2}}}~,~~~~~\beta=\dfrac{1}{\sqrt{ 4v_{\Phi_{1}}^{2}v_{\Phi_{2}}^{2}+v_{\Phi_{1}}^{2}v_{s}^{2}+v_{\Phi_{2}}^{2}v_{s}^{2} }}~.
\end{equation}
Unless $\kappa$ is tiny, neither the CP-even (squared) masses $m_{H_{k}}^{2}$ nor $\mathcal{O}_{R}$ have a simple analytical form, they are numerically evaluated instead. Turning now to the charged sector we have, in the basis $(\varphi_{1}^{+},\varphi_{2}^{+})$, the following mass squared matrix
 \begin{eqnarray}
 M_{\pm}^2=\dfrac{1}{2}\left(
\begin{array}{cc}
\left(-\lambda'_{12} v_{\Phi_2}  + \frac{\kappa  v_s^2}{v_{\Phi_1}}\right) v_{\Phi_2} & \lambda'_{12} v_{\Phi_1} v_{\Phi_2} -\kappa  v_s^2  \\
 \lambda'_{12} v_{\Phi_1} v_{\Phi_2} -\kappa  v_s^2  &  \left(-\lambda'_{12} v_{\Phi_1} +\frac{\kappa  v_s^2}{v_{\Phi_2} }\right)v_{\Phi_1}~ \\
\end{array}\right)~,
\end{eqnarray}
whose eigenstates are the longitudinal $W^{\pm}$ boson and the physical $H^{\pm}$. The (squared) masses and rotation matrix satisfying $\text{diag}(0,m_{H_{\pm}}^{2})=\mathcal{O}_{\pm}M_{\pm}^{2}\mathcal{O}_{\pm}^{T}$ are
\begin{equation}
m_{H^{\pm}}^2= \dfrac{1}{2}\left(v_{\Phi_1}^2+v_{\Phi_2} ^2\right) \left(-\lambda'_{12}+\frac{\kappa  v_s^2}{v_{\Phi_1} v_{\Phi_2} }\right)~,~~~~~
\mathcal{O}_{\pm}=\left(\begin{array}{cc}
 \alpha v_{\Phi_{1}} & \alpha v_{\Phi_{2}} \\
-\alpha v_{\Phi_{2}} & \alpha v_{\Phi_{1}}
\end{array}\right)~.
\end{equation}

\subsection{Charged fermion sector}

From Eq. (\ref{eq:LYukawa}), the textures of the charged fermion mass matrices are

\begin{eqnarray}
 M_\ell=\left(\begin{array}{ccc}
 y_{1e} v_{\Phi_1}  & y_{1\mu} v_{\Phi_2}   & y_{1\tau} v_{\Phi_1} \\
 y_{2e} v_{\Phi_1}  & y_{2\mu} v_{\Phi_2}   & y_{2\tau} v_{\Phi_1}  \\
 y_{3e} v_{\Phi_1}  & y_{3\mu} v_{\Phi_2}   & y_{3\tau} v_{\Phi_1}\\
\end{array}\right)~,
\end{eqnarray}

\begin{eqnarray}
 M_{d}=\left(\begin{array}{ccc}
 y_{1d} v_{\Phi_2}  & y_{1s} v_{\Phi_1}   & y_{1b} v_{\Phi_1} \\
 y_{2d} v_{\Phi_2}  & y_{2s} v_{\Phi_1}   & y_{2b} v_{\Phi_1}  \\
 y_{3d} v_{\Phi_2}  & y_{3s} v_{\Phi_1}   & y_{3b} v_{\Phi_1}\\
\end{array}\right)~,
\end{eqnarray}

\begin{eqnarray}
 M_{u}=\left(\begin{array}{ccc}
 y_{1u} v_{\Phi_2}  & y_{1c} v_{\Phi_1}   & y_{1t} v_{\Phi_1} \\
 y_{2u} v_{\Phi_2}  & y_{2c} v_{\Phi_1}   & y_{2t} v_{\Phi_1}  \\
 y_{3u} v_{\Phi_2}  & y_{3c} v_{\Phi_1}   & y_{3t} v_{\Phi_1}\\
\end{array}\right)~.
\end{eqnarray}
Given the number of Yukawas in the up and down quark mass matrices, the Cabbibo-Kobayashi-Maskawa mixing matrix can be straightforwardly fitted.

\section{$\boldsymbol{\Delta a_{\mu}}$ contributions}
\label{sec:appB}

The notation and normalization for the muon couplings in Secs. \ref{sec:radius} and \ref{sec:gminus2} is adopted from Ref. \cite{Lindner:2016bgg}, with $S$, $P$ denoting true scalar and pseudoscalar, $V$ and $A$ polar and axial vector, and $\widetilde{S}, \widetilde{P}$ refering to charged scalar couplings
\begin{align}
\mathcal{L}
&\supset~C_{S}^{ij}~\phi^{0}\overline{\ell_{i}}\ell_{j}~+~C_{P}^{ij}~\phi^{0}\overline{\ell_{i}}\gamma_{5}\ell_{j} \notag \\
&+ \biggl( \widetilde{C}_{S}^{ij}~\phi^{+}\overline{\nu}_{i}\ell_{j}~+~\widetilde{C}_{P}^{ij}~\phi^{+}\overline{\nu}_{i}\gamma_{5}\ell_{j}~+\text{H.c.} \biggr) \notag \\
&+ \biggl( C_{V}^{ij}Z'_{\mu}\overline{\ell}_{i}\gamma^{\mu}\ell_{j}~+C_{A}^{ij}Z'_{\mu}\overline{\ell}_{i}\gamma^{\mu}\gamma^{5}\ell_{j}~+\text{H.c.} \biggr)~.
\end{align}
With vanishing charged lepton mixing, for $i=j=\mu$
\begin{equation}
\Delta a_{\mu}^{(Z')}=\dfrac{1}{8\pi^{2}}\dfrac{m_{\mu}^{2}}{m_{Z'}^{2}}\int_{0}^{1}\text{d}x~\dfrac{\bigl| C_{V}[Z'] \bigr|^{2}P_{V}(x)+\bigl| C_{A}[Z'] \bigr|^{2}P_{A}(x)}{(1-x)(1-m_{\mu}^{2}/m_{Z'}^{2})+x(m_{\mu}^{2}/m_{Z'}^{2})} \label{eq:DeltaAmuZp}
\end{equation}
where
\begin{equation}
P_{V,A}(x)\equiv 2x(1-x)(x-2\pm2)+(m_{\mu}^{2}/m_{Z'}^{2})x^{2}(1\mp1)^{2}(1-x\pm1)~.
\end{equation}
For the $CP$-even scalars $H_{k}$~,
\begin{equation}
\Delta a_{\mu}^{(H_{k})}=\dfrac{1}{8\pi^{2}}\dfrac{m_{\mu}^{2}}{m_{H_{k}}^{2}}\int_{0}^{1}\text{d}x~\dfrac{\bigl| C^{\mu}_{S}[H_{k}] \bigr|^{2}P_{S}(x)}{(1-x)(1-m_{\mu}^{2}/m_{H_{k}}^{2})+x(m_{\mu}^{2}/m_{H_{k}}^{2})} \label{eq:DeltaAmuS}
\end{equation}
with
\begin{equation}
P_{S}(x)\equiv x^{2}(2-x)~.
\end{equation}
The analog pseudoscalar and charged scalar contributions are listed below,
\begin{equation}
\Delta a_{\mu}^{(A)}=\dfrac{1}{8\pi^{2}}\dfrac{m_{\mu}^{2}}{m_{A}^{2}}\int_{0}^{1}\text{d}x~\dfrac{\bigl| C^{\mu}_{P}[A] \bigr|^{2}P_{P}(x)}{(1-x)(1-m_{\mu}^{2}/m_{A}^{2})+x(m_{\mu}^{2}/m_{A}^{2})}
\end{equation}
with
\begin{equation}
P_{P}(x)\equiv -x~,
\end{equation}
and
\begin{equation}
\Delta a_{\mu}^{(H^{+})}=\dfrac{1}{8\pi^{2}}\dfrac{m_{\mu}^{2}}{m_{H^{+}}^{2}}\int_{0}^{1}\text{d}x~\dfrac{\bigl| \widetilde{C}^{\mu}_{S}[H^{+}] \bigr|^{2}P_{\widetilde{S}}(x) + \bigl| \widetilde{C}^{\mu}_{P}[H^{+}] \bigr|^{2}P_{\widetilde{P}}(x)}{(1-x)(1-m_{\mu}^{2}/m_{H^{+}}^{2})+x(m_{\mu}^{2}/m_{H^{+}}^{2})}
\end{equation}
where
\begin{equation}
P_{\widetilde{S},\widetilde{P}}(x)\equiv x(1-x)(x\pm1)~.
\end{equation}

\bibliographystyle{utphys}
\bibliography{thebiblio}

\end{document}